\documentclass[preprint,10pt]{popl14conf} 

\newcommand{\xtitle}{Adaptive Interpreter Optimization using Multi-Level Quickening}
\renewcommand{\xtitle}{Speculative Staging for Interpreter Optimization}

\usepackage{amsmath}
\usepackage{amsthm}             
\usepackage{stmaryrd}           
\usepackage{fancyhdr}
\usepackage{subfig}             
\usepackage[usenames,svgnames,table]{xcolor}
\usepackage{graphicx}           
\usepackage{listings}           
\usepackage{multicol}
\usepackage{multirow}
\usepackage{verbatim}
\usepackage{enumerate}          
\usepackage{underscore}         
\usepackage{todonotes}          
\usepackage[ruled]{algorithm2e} 
\usepackage{eufrak}             
\usepackage{fixltx2e}           
\usepackage{tikz}               
\usepackage{microtype}          
\usepackage[pdftex,
            pdfauthor={stefan brunthaler},
            pdftitle={Speculative Staged Interpreter Optimization},
            pdfsubject={Speculative Staged Interpreter Optimization},
            pdfkeywords={Interpreter, Inline Caching, Type Feedback, Type Propagation, Quickening, Optimization},
            letterpaper]{hyperref}

\usetikzlibrary{arrows,chains,calc,matrix,positioning,scopes,shapes,decorations.pathreplacing,shapes.arrows,shapes.multipart,patterns}

\def\algorithmautorefname{Algorithm}

\lstdefinestyle{prettyhaskell}{
  language=Haskell,
  basicstyle=\small\ttfamily,
  flexiblecolumns=false,
  numbersep=5pt,
  tabsize=2,
  extendedchars=true,
  keywordstyle=\color{blue}\bfseries,
  commentstyle=\color{red}\itshape,
  basewidth={0.5em,0.45em},
  numbers=left,
  numberstyle=\tiny,
  mathescape=true,
  literate={+}{{$+$}}1 {/}{{$/$}}1 {*}{{$*$}}1 {=}{{$=$}}1
           {>}{{$>$}}1 {<}{{$<$}}1 {\\}{{$\lambda$}}1
           {\\\\}{{\char`\\\char`\\}}1
           {->}{{$\rightarrow$}}2 {>=}{{$\geq$}}2 {<-}{{$\leftarrow$}}2
           {<=}{{$\leq$}}2 {=>}{{$\Rightarrow$}}2
           {\ .}{{$\circ$}}2 {\ .\ }{{$\circ$}}2
           {>>}{{>>}}2 {>>=}{{>>=}}2
           {|}{{$\mid$}}1
}

\lstdefinestyle{othercode}{
  basicstyle=\scriptsize\ttfamily, 
  numbers=left,               
  numberstyle=\tiny,          
  numbersep=5pt,              
  tabsize=2,                  
  extendedchars=true,         %
  keywordstyle=\color{blue}\bfseries,
  commentstyle=\color{red}\itshape,
  stringstyle=\color{Green}\ttfamily, 
  showspaces=false,           
  showtabs=false,             
  showstringspaces=false      
}

\newcommand{\sbr}[1]{}

\newcommand{\coldata}[1]{\texttt{#1}}

\newcommand{\pyType}[1]{\texttt{Py#1\_Type}}

\newcommand{\namasteMaxSU}{4.222}
\newcommand{\incaMaxSU}{1.7362}

\newcommand{\fasterThanINCA}{143\%}

\newcommand{\namaste}{\texttt{NAMA}}
\newcommand{\inca}{\texttt{INCA}}
\newcommand{\mlq}{\texttt{MLQ}}

\makeatletter
\let\@copyrightspace\relax
\makeatother

\begin{document}


\title{\xtitle}


\authorinfo{Stefan Brunthaler}
           {University of California, Irvine}
           {s.brunthaler@uci.edu}





\maketitle

\begin{abstract}
Interpreters have a bad reputation for having lower performance than just-in-time compilers.
We present a new way of building high performance interpreters that is particularly effective for
executing dynamically typed programming languages.
The key idea is to combine speculative staging of optimized interpreter instructions with a novel
technique of incrementally and iteratively concerting them at run-time.

This paper introduces the concepts behind deriving optimized instructions from existing interpreter
instructions---incrementally peeling off layers of complexity.
When compiling the interpreter, these optimized derivatives will be compiled along with the original
interpreter instructions.
Therefore, our technique is portable by construction since it leverages the existing compiler's
backend.
At run-time we use instruction substitution from the interpreter's original and expensive
instructions to optimized instruction derivatives to speed up execution.

Our technique unites high performance with the simplicity and portability of interpreters---we
report that our optimization makes the CPython interpreter up to more than four times faster, where
our interpreter closes the gap between and sometimes even outperforms PyPy's just-in-time compiler.
\end{abstract}


\terms{Design, Languages, Performance}



\section{Introduction}\label{s:intro}

The problem with interpreters for dynamically typed programming languages is that they are slow.
The fundamental lack of performance is due to following reasons.
First, their implementation is simple and does not perform known interpreter optimizations, such
as threaded code~\cite{bell+73,debaere.campenhout+90,berndl.etal+05} or
superinstructions~\cite{proebsting+95,ertl.gregg+03,ertl.gregg+04}.
Second, even if the interpreters apply these known techniques, their performance potential is
severely constrained by expensive interpreter instruction implementations~\cite{brunthaler+09}.

Unfortunately, performance-conscious implementers suffer from having only a limited set of options
at their disposal to solve this problem.
For peak performance, the current best-practice is to leverage results from dynamic compilation.
But, implementing a just-in-time, or JIT, compiler is riddled with many problems, e.g., a lot of
tricky details, hard to debug, and substantial implementation effort.
An alternative route is to explore the area of purely interpretative optimization instead.
These are optimizations that preserve innate interpreter characteristics, such as
ease-of-implementation and portability, while offering important speedups.
Prior work in this area already reports the potential of doubling the execution
performance~\cite{brunthaler+10a,brunthaler+10b}.
As a result, investigating a general and principled strategy for optimizing high abstraction-level
interpreters is particularly warranted.

Interpreting a dynamically typed programming language, such as JavaScript, Python, or Ruby, has its
own challenges.
Frequently, these interpreters use one or a combination of the following features:
\begin{itemize}
\item dynamic typing to select type-specific operations,
\item reference counting for memory management, and
\item modifying boxed data object representations.
\end{itemize}
To cope with these features, interpreter instructions naturally become expensive in terms of
assembly instructions required to implement their semantics.
Looking at successful research in just-in-time compilation, we know that in order to achieve
substantial performance improvements, we need to reduce the complexity of the interpreter
instructions' implementation.
Put differently, we need to remove the overhead introduced by dynamic typing, reference counting,
and operating on boxed data objects.

In this paper we combine ideas from staged compilation with partial evaluation and interpreter
optimization to devise a general framework for interpreter optimization.
From staged compilation, we take the idea that optimizations can spread out across several stages.
From partial evaluation, we take inspiration from the Futamura projections to optimize programs.
From interpreter optimization, we model a general theory of continuous optimization based on
rewriting instructions.
These ideas form the core of our framework, which is purely interpretative, i.e., it offers
ease-of-implementation and portability while avoiding dynamic code generation, and delivers high
performance.
As a result, close to three decades after Deutsch and Schiffman~\cite{deutsch.schiffman+84}
described the ideas of what would eventually become the major field of just-in-time compilation, our
framework presents itself as an alternative to implementing JIT compilers.

Summing up, this paper makes the following contributions.
\begin{itemize}
\item We introduce a general theory for optimizing interpreter instructions that relies on
  speculatively staging of optimized interpreter instructions at interpreter compile-time and
  subsequent concerting at run-time (see~\autoref{s:theory}).

\item We present a principled procedure to derive optimized interpreter instructions via partial
  evaluation.
  Here, speculation allows us to remove previous approaches' requirement to specialize
  towards a specific program (see~\autoref{ss:theory}).

\item We apply the general theory to the Python interpreter and describe the relevant implementation
  details (see~\autoref{s:implementation}).

\item We provide results of a careful and detailed evaluation of our Python based implementation
  (see~\autoref{s:evaluation}), and report substantial speedups of up to more than four times faster
  than the CPython implementation for the \textsf{spectralnorm} benchmark.
  For this benchmark our technique outperforms the current state-of-the-art JIT compiler, PyPy 1.9,
  by 14\%.
\end{itemize}

\section{Example}\label{s:example}

In this section we walk through a simple example that illustrates how interpreters address---or
rather fail to address---the challenge of efficiently executing a high-level language.
The following listing shows a Python function \texttt{sum} that ``adds'' its parameters and returns
the result of this operation.
\begin{lstlisting}[language=Python,morekeywords={JUMPTO,DISPATCH,r13},numbers=left,numberstyle=\tiny,style=othercode]
def sum(a, b):
    return a + b
\end{lstlisting}

In fact, this code does not merely ``add'' its operands: depending on the actual types of the
parameters \texttt{a} and \texttt{b}, the interpreter will select a matching operation.
In Python, this means that it will either concatenate strings, or perform arithmetic addition on
either integers, floating point numbers, or complex numbers; or the interpreter could even invoke
custom Python code---which is possible due to Python's support for ad-hoc polymorphism.

In 1984, Deutsch and Schiffman~\cite{deutsch.schiffman+84} report that there exists a ``dynamic
locality of type usage,'' which enables speculative optimization of code for any arbitrary but fixed
and observed type $\tau$.
Subsequent research into dynamic compilation capitalizes on this observed locality by speculatively
optimizing code using \emph{type feedback}~\cite{holzle.ungar+94,holzle+94}.
From their very beginning, these dynamic compilers---or just-in-time compilers as they are referred
to frequently---had to operate within a superimposed latency time.
Put differently, dynamic compilers traditionally sacrifice known complex optimizations for
predictable compilation times.

Staged compilation provides a solution to the latency problem of JIT compilers by distributing work
needed to assemble optimized code among separate stages.
For example, a staged compiler might break up an optimization to perform work at compile time,
link-time, load-time, or finally at run-time.
The problem with staged optimizations for high-level languages such as JavaScript, Python, and Ruby
is that they require at least partial knowledge about the program.
But, as the example of the \texttt{sum} function illustrates, only at run-time we will actually
identify the concrete type $\tau$ for parameters \texttt{a} and \texttt{b}.

For our target high-level languages and their interpreters, staged compilation is not possible,
primarily due to two reasons.
First, none of these interpreters have a JIT compiler, i.e., preventing staged partial
optimizations.
Second, the stages of staged compilation and interpreters are separate.
The traditional stages listed above need to be partitioned into stages that we need to assemble the
interpreter (viz.~compile-time, link-time, and load-time), and separate stages of running the
program: at interpreter run-time, it compiles, potentially links, loads and runs hosted programs.

\section{Speculative Staged Interpreter Optimization}\label{s:theory}

The previous section sketches the problem of performing traditional staged optimizations for
interpreters.
In this section we are first going to dissect interpreter performance to identify bottlenecks.
Next, we are going to describe which steps are necessary to formalize the problem, and subsequently
use speculative staged interpreter optimizations to conquer them and achieve high performance.

\subsection{Dissecting Example Performance Obstacles}\label{ss:dissecting-perf}

\newcommand{\pySum}{\texttt{sum}}
\newcommand{\loadFast}{\texttt{LOAD_FAST}}
\newcommand{\storeFast}{\texttt{STORE_FAST}}
\newcommand{\binaryAdd}{\texttt{BINARY_ADD}}

Python's compiler will emit the following sequence of interpreter instructions, often called
bytecodes, when compiling the \pySum{} function (ignoring argument bytes for the \loadFast{}
instructions):
\begin{center}
  \begin{tikzpicture}[scale=.75, every node/.style={scale=.75}]
    \node (l1) [draw,rectangle,minimum width=1.25cm,minimum height=1.25cm,align=center] {\texttt{LOAD}\\\texttt{FAST}};
    \node (l2) at ($(l1.east)+(1pt,0)$) [draw,rectangle,minimum width=1.25cm,minimum height=1.25cm,anchor=west, align=center] {\texttt{LOAD}\\\texttt{FAST}};
    \node (add) at ($(l2.east)+(1pt,0)$) [draw,rectangle,minimum width=1.25cm,minimum height=1.25cm,anchor=west, align=center] {\texttt{BINARY}\\\texttt{ADD}};
    \node (ret) at ($(add.east)+(1pt,0)$) [draw,rectangle,minimum width=1.25cm,minimum height=1.25cm,anchor=west, align=center] {\texttt{RETURN}\\\texttt{VALUE}};
  \end{tikzpicture}
\end{center}

We see that the interpreter emits untyped, polymorphic instructions that rely on dynamic typing to
actually select the operation.
Furthermore, we see that Python's virtual machine interpreter implements a stack to pass operand
data between instructions.

Let us consider the application \texttt{sum(3, 4)}, i.e., \texttt{sum} is used with the specific
type $\mathtt{Int} \times \mathtt{Int} \rightarrow \mathtt{Int}$.
In this case, the \binaryAdd{} instruction will check operand types and select the proper operation
for the types.
More precisely, assuming absence of ad-hoc polymorphism, the Python interpreter will identify that
both \texttt{Int} operands are represented by a C \texttt{struct} called \pyType{Long}.
Next, the interpreter will determine that the operation to invoke is
\pyType{Long}\texttt{->tp_as_number->nb_add}, which points to the \texttt{long_add} function.
This operation implementation function will then unbox operand data, perform the actual integer
arithmetic addition, and box the result.
In addition, necessary reference counting operations enclose these operations, i.e., we need to
decrease the reference count of the arguments, and increase the reference count of the result.
Both, (un-)boxing and adjusting reference count operations add to the execution overhead of the
interpreter.

Contrary to the interpreter, a state-of-the-art JIT compiler would generate something like this
\begin{lstlisting}[morekeywords={rax,rbx,rsp,movq,addq,ret},numbers=left,numberstyle=\tiny,style=othercode]
movq %rax, -8(%rsp)
movq %rbx, -16(%rsp)
addq %rax, %rbx
ret
\end{lstlisting}
The first two lines assume a certain stack layout to where to find operands \texttt{a} and
\texttt{b}, both of which we assume to be unboxed.
Hence, we can use the native machine addition operation (line 3) to perform arithmetic addition and
return the operation result in \texttt{\%rax}.

Bridging the gap between the high abstraction-level representation of computation in Python
bytecodes and the low abstraction-level representation of native machine assembly instructions holds
the key for improving interpreter performance.
To that end, we classify both separate instruction sets accordingly:
\begin{itemize}
\item Python's instruction set is untyped and operates exclusively on boxed objects.
\item Native-machine assembly instructions are typed and directly modify native machine data.
\end{itemize}

An efficient, low-level interpreter instruction set allows us to represent the \pySum{} function's
computation for our assumed type in the following way:
\begin{center}
  \begin{tikzpicture}[scale=.75, every node/.style={scale=.75}]
    \node (l1) [draw,rectangle,minimum width=1.25cm,minimum height=1.25cm,align=center] {\texttt{LOAD}\\\texttt{INT}};
    \node (l2) at ($(l1.east)+(1pt,0)$) [draw,rectangle,minimum width=1.25cm,minimum height=1.25cm,anchor=west, align=center] {\texttt{LOAD}\\\texttt{INT}};
    \node (add) at ($(l2.east)+(1pt,0)$) [draw,rectangle,minimum width=1.25cm,minimum height=1.25cm,anchor=west, align=center] {\texttt{INT}\\\texttt{ADD}};
    \node (ret) at ($(add.east)+(1pt,0)$) [draw,rectangle,minimum width=1.25cm,minimum height=1.25cm,anchor=west, align=center] {\texttt{RETURN}\\\texttt{INT}};
  \end{tikzpicture}
\end{center}

In this lower-level instruction set the instructions are typed, which allows using a different
operand data passing convention, and directly modifying unboxed data---essentially operating at the
same semantic level as the assembly instructions shown above; disregarding the different
architectures, i.e., register vs.~stack.

\subsection{Systematic Derivation}\label{ss:theory}

The previous section informally discusses the goal of our speculatively staged interpreter
optimization: optimizing from high to low abstraction-level instructions.
This section systematically derives required components for enabling interpreters to use this
optimization technique.
In contrast with staged compilation, staged interpreter optimization is purely interpretative, i.e.,
it avoids dynamic code generation altogether.
The key idea is that we:
\begin{itemize}
\item stage and compile new interpreter instructions at interpreter compile-time,
\item concert optimized instruction sequences at run-time by the interpreter.
\end{itemize}

The staging process involves the ahead-of-time compiler that is used to compile the interpreter.
Therefore, this process exercises the compiler's backend to have portable code generation and
furthermore allows the interpreter implementation to remain simple.
The whole process is, however, speculative:
only by actually interpreting a program, we know for sure which optimized interpreter instructions
are required.
As a result, we restrict ourselves to generate optimized code only for instructions that have a high
likelihood of being used.

To assemble optimized instruction sequences at run-time, we rely on a technique known as
\emph{quickening}~\cite{lindholm.yellin+96}.
Quickening means that we replace instructions with optimized derivatives of the exact same
instruction at run-time.
Prior work focuses on using only one level of quickening, i.e., replacing one instruction with
another instruction derivative.
In this work, we introduce \emph{multi-level quickening}, i.e., the process of iteratively and
incrementally rewriting interpreter instructions to ever more specialized derivatives of the generic
instruction.

\subsubsection{Prerequisites}

In this section we present a simplified substrate of a dynamically typed programming language
interpreter, where we illustrate each of the required optimization steps.
\begin{lstlisting}[style=prettyhaskell]
data Value = VBool Bool
           | VInt Int
           | VFloat Float
           | VString String

type Stack = [Value]
type Instr = Stack -> Stack
\end{lstlisting}

We use \texttt{Value} to model a small set of types for our interpreter.
The operand stack \texttt{Stack} holds intermediate values, and an instruction \texttt{Instr} is a
function modifying the operand stack \texttt{Stack}.
Consequently, the following implementation of the interpreter keeps evaluating instructions until
the list of instructions is empty, whereupon it returns the top of the operand stack element as its
result:
\begin{lstlisting}[style=prettyhaskell]
interp :: Stack -> [Instr] -> Value
interp (x:xs) [] = x
interp s (i:is)  =
  let
    s'= i s
  in
    eval s' is
\end{lstlisting}

Using \texttt{interp} as the interpreter, we can turn our attention to the actual interpreter
instructions.
The following example illustrates a generic implementation of a binary interpreter instruction,
which we can instantiate, for example for an arithmetic add operation:

\begin{lstlisting}[style=prettyhaskell]
binaryOp :: (Value -> Value -> Value) -> Stack -> Stack
binaryOp f s =
  let
    (x, s') = pop s
    (y, s'') = pop s'
  in
   (f x y):s''

addOp :: Stack -> Stack
addOp = binaryOp dtAdd

dtAdd :: Value -> Value -> Value
dtAdd (VInt x) (VInt y) = (VInt (x + y))
dtAdd (VFloat x) (VFloat y) = (VFloat (x + y))
dtAdd (VString x) (VString y) = (VString (x ++ y))
dtAdd (VBool x) (VBool y) = (VBool (x && y))
$\ldots$
\end{lstlisting}

The generic implementation \texttt{binaryOp} shows the operand stack modifications all binary
operations need to follow.
The \texttt{addOp} function implements the actual resolving logic of the dynamic types of
\texttt{Value}s via pattern matching starting on line 13 in function \texttt{dtAdd}.

\subsubsection{First-level Quickening: Type Feedback}

\theoremstyle{definition}
\newtheorem{defn}{Definition}
\newtheorem{exmp}{Example}

\begin{defn}[Instruction Derivative]
  An instruction $I'$ is an \emph{instruction derivative} of an instruction $I$, if and only if it
  implements the identical semantics for an arbitrary but fixed subset of $I$'s functionality.
\end{defn}

\begin{exmp}
  In our example interpreter \texttt{interp}, the \texttt{addOp} instruction has type
  $\mathtt{Value} \times \mathtt{Value} \rightarrow \mathtt{Value}$.
  An instruction derivative \texttt{intAdd} would implement the subset case of integer arithmetic
  addition only, i.e., where operands have type \texttt{VInt}.
  Analogous cases are for all combinations of possible types, e.g., for string concatenation
  (\texttt{VString}).
\end{exmp}

To obtain all possible instruction derivatives $I'$ for any given interpreter instruction $I$, we
rely on insights obtained by partial evaluation~\cite{jones.etal+93}.
The first Futamura projection~\cite{futamura+71} states that we can derive a compiled version of an
interpreted program $\Pi$ by partially evaluating its corresponding interpreter \texttt{interp}
written in language~$L$:
\begin{equation}
  \label{eq:first-futamura}
  \mathtt{compiled}\Pi := \llbracket \mathtt{mix} \rrbracket_{L} [ \mathtt{interp}, \Pi ]
\end{equation}
In our scenario, this is not particularly helpful, because we do not know program $\Pi$ a priori.
The second Futamura projection tells us how to derive a compiler by applying \texttt{mix} to itself
with the interpreter as its input:
\begin{equation}
  \label{eq:second-futamura}
  \mathtt{compiler} := \llbracket \mathtt{mix} \rrbracket_{L} [ \mathtt{mix}, \mathtt{interp} ]
\end{equation}
By using the interpreter \texttt{interp} as its input program, the second Futamura projection
eliminates the dependency on the input program~$\Pi$.
However, for a dynamically typed programming language, a compiler derived by applying the second
Futamura projection without further optimizations is unlikely to emit efficient code---because it
lacks information on the types~\cite{thibault.etal+00}.

Our idea is to combine these two Futamura projections in a novel way:
\begin{equation}
  \label{eq:deriving-instrs}
  \displaystyle\mathop{\mathlarger{\mathlarger{\mathlarger{\forall}}}}_{I \in \mathtt{interp}}
  I_{\tau}' := \llbracket \mathtt{mix} \rrbracket_{L} [ I, \tau ]
\end{equation}
That means that for all instructions $I$ of an interpreter \texttt{interp}, we derive an optimized
instruction derivative $I'$ specialized to a type $\tau$ by partially evaluating an instruction $I$
with the type $\tau$.
Hence, we speculate on the likelihood of the interpreter operating on data of type $\tau$ but do
eliminate the need to have a priori knowledge about the program $\Pi$.

To preserve semantics, we need to add a guard statement to $I'$ that ensures that the actual operand
types match up with the specialized ones.
If the operand types do not match, we need to take corrective measures and redirect control back to
$I$.
For example, we get the optimized derivative \texttt{intAdd} from \texttt{dtAdd} by fixing the
operand type to \texttt{VInt}:
\begin{equation}
  \label{eq:deriving-instrs}
  \mathtt{intAdd} := \llbracket \mathtt{mix} \rrbracket_{L} [ \mathtt{dtAdd}, \mathtt{VInt} ]
\end{equation}
\begin{lstlisting}[style=prettyhaskell]
intAdd :: Value -> Value -> Value
intAdd (VInt x) (VInt y) = (VInt (x + y))
intAdd x y = dtAdd x y
\end{lstlisting}

The last line in our code example acts as a guard statement, since it enables the interpreter to
execute the type-generic \texttt{dtAdd} instruction whenever the speculation of \texttt{intAdd} fails.

The interpreter can now capitalize on the ``dynamic locality of type
usage''~\cite{deutsch.schiffman+84} and speculatively eliminate the overhead of dynamic typing by
rewriting an instruction $I$ at position $p$ of a program $\Pi$ to its optimized derivative $I'$:
\begin{equation}
  \label{eq:quickening}
  (\Pi)[I'/ I_{p}]
\end{equation}

It is worth noting that this rewriting, or \emph{quickening} as it is commonly known, is purely
interpretative, i.e., it does not require any dynamic code generation---simply updating the
interpreted program suffices:

\begin{lstlisting}[style=prettyhaskell]
quicken :: [ Instr ] -> Int -> Instr -> [ Instr ]
quicken $\Pi$ p derivative =
  let
    (x, y:ys)= splitAt p $\Pi$
    r= x ++ derivative : ys
\end{lstlisting}

Using this derivation step, we effectively create a typed interpreter instruction set for an
instruction set originally only consisting of untyped interpreter instructions.

\subsubsection{Second-level Quickening: Type Propagation}

Taking a second look at the optimized derivative instruction \texttt{intAdd} shows that it still
contains residual overhead: unboxing the \texttt{VInt} operands and boxing the \texttt{VInt} result
(cf.~line three).
It turns out that we can do substantially better by identifying complete expressions that operate on
the same type and subsequently optimizing the whole sequence.

For identifying a sequence of instructions that operate on the same type, we leverage the type
information collected at run-time via the first-level quickening step described in the previous
section.
During interpretation, the interpreter will optimize programs it executes on the fly and single
instruction occurrences will carry type information.
To expand this information to bigger sequences, we use an abstract interpreter that propagates
the collected type information.
Once we have identified a complete sequence of instructions operating on the same type, we then
quicken the complete sequence to another set of optimized derivatives that directly modify unboxed
data.
Since these instructions operate directly on unboxed data, we need to take care of several pending
issues.

First, operating on unboxed data requires modifying the operand stack data passing convention.
The original instruction set, as well as the optimized typed instruction set, operates on boxed
objects, i.e., all elements on the operand stack are just pointers to the heap, having the type of
one machine word (\texttt{uint64} on modern 64-bit architectures).
If we use unboxed data elements, such as native machine integers and floats, we need to ensure that
all instructions follow the same operand passing convention.

\begin{defn}[Operand Stack Passing Convention]
  All instructions operating on untyped data need to follow the same operand stack data passing
  convention.
  Therefore, we define a conversion function~$\mathfrak{c}$ to map data to and from at least one
  native machine word:
  \begin{eqnarray}
    \label{eq:stack-conversion}
    \mathfrak{c}_{\tau}: \tau \rightarrow \mathtt{uint64}^{+}\\
    \mathfrak{c}_{\tau}^{-1}: \mathtt{uint64}^{+} \rightarrow \tau
  \end{eqnarray}
\end{defn}

Second, we need to provide and specify dedicated (un-)boxing operations to unbox data upon entering a
sequence of optimized interpreter instructions, and box results when leaving an optimized sequence.
\begin{defn}[Boxing and Unboxing of Objects]
  We define a function $\mathfrak{m}$ to map objects to at least one machine word and conversely
  from at least one machine word back to proper language objects of type $\pi$.
  \begin{eqnarray}
    \label{eq:stack-conversion}
    \mathfrak{m}_{\pi}: \pi \rightarrow \tau^{+}\\
    \mathfrak{m}_{\pi}^{-1}: \tau^{+} \rightarrow \pi
  \end{eqnarray}
\end{defn}
Third, this optimization is speculative, i.e., we need to take precautions to preserve semantics
and recover from misspeculation.
Preserving semantics of operating on unboxed data usually requires to use a tagged data format
representation, where we reserve a set of bits to hold type information.
But, we restrict ourselves to sequences where we know the types \emph{a priori}, which allows us to
remove the restrictions imposed by using a tagged data format, i.e., additional checking code and
decreasing range of representable data.
In general, whenever our speculation fails we need to generalize specialized instruction
occurrences back to their more generic instructions and resume regular interpretation.

In all definitions, we use the $^{+}$ notation to indicate that a concrete instantiation of either
$\mathfrak{c}$ or $\mathfrak{m}$ is able to project data onto multiple native machine words.
For example, the following implementation section will detail one such case where we represent
Python's complex numbers by two native machine words.

\paragraph{Abstract Interpretation}

Taking inspiration from Leroy's description of Java bytecode verification~\cite{leroy+03}, we also
use an abstract interpreter that operates over types instead of values.
Using the type information captured in the previous step, for example from quickening from the type
generic \texttt{addOp} to the optimized instruction derivative \texttt{intAdd}, we can propagate
type information from operation instructions to the instructions generating its operands.

For example, we know that the \texttt{intAdd} instruction expects its operands to have type
\texttt{Int} and produces an operand of type \texttt{Int}:
\begin{equation*}
  \mathtt{intAdd} : (\mathtt{VInt}.\mathtt{VInt}.S) \rightarrow (\mathtt{VInt}.S)
\end{equation*}
Similar to our actual interpreter, the abstract interpreter uses a transition relation $i: S
\rightarrow S'$
to represent an instruction $i$'s effect on the operand stack.
All interpreter instructions of the original instruction set are denoted by type-generic rules that
correspond to the top element of the type lattice, i.e., in our case \texttt{Value}.
The following rules exemplify the representation, where we only separate instructions by their
arity:
\begin{align*}
  \mathtt{unaryOp}  &: (\mathtt{Value}.S) \rightarrow (\mathtt{Value}.S) \\
  \mathtt{binaryOp} &: (\mathtt{Value}.\mathtt{Value}.S) \rightarrow (\mathtt{Value}.S)
\end{align*}

The set of types our abstract interpreter operates on corresponds to the set of types we generated
instruction derivatives for in the first-level quickening step, i.e., (\texttt{Int}, \texttt{Bool},
\texttt{Float}, \texttt{String}).
For simplicity, our abstract interpreter ignores branches in the code, which limits our scope to
propagate types along straight-line code, but on the other hand avoids data-flow analysis and
requires only one linear pass to complete.
This is important insofar as we perform this abstract interpretation at run-time and therefore are
interested to keep latency introduced by this step at a minimum.

Type propagation proceeds as follows.
The following example shows an original example program representation as emitted by some other
program, e.g., another compiler:
\begin{equation*}
  [\cdots, \mathtt{push}^{0}, \mathtt{push}^{1}, \mathtt{addOp}^{2}, \mathtt{push}^{3}, \mathtt{addOp}^{4}, \mathtt{pop}^{5}, \cdots]
\end{equation*}
After executing this example program, the first-level quickening captures types encountered during
execution:
\begin{equation*}
  [\cdots, \mathtt{push}^{0}, \mathtt{push}^{1}, \mathtt{intAdd}^{2}, \mathtt{push}^{3}, \mathtt{intAdd}^{4}, \mathtt{pop}^{5}, \cdots]
\end{equation*}
Now, we propagate the type information by abstract interpretation.
Since \texttt{intAdd} expects operands of type \texttt{Int}, we can infer that the first two
\texttt{push} instructions must push operands of type \texttt{Int} onto the operand stack.
Analogously, the second occurrence of \texttt{intAdd} allows us to infer that the result of the
first \texttt{intAdd} has type \texttt{Int}, as does the third occurrence of the \texttt{push}
instruction.
Finally, by inspecting the type stack when the abstract interpreter reaches the \texttt{pop}
instruction, we know that it must pop an operand of type \texttt{Int} off the stack.
Therefore, after type propagation our abstract interpreter will have identified that the complete
sequence of instructions actually operate exclusively on data of type \texttt{Int}:
\begin{equation*}
  [\cdots, \mathtt{push}_{\mathtt{Int}}^{0}, \mathtt{push}_{\mathtt{Int}}^{1}, \mathtt{intAdd}^{2}, \mathtt{push}_{\mathtt{Int}}^{3}, \mathtt{intAdd}^{4},  \mathtt{pop}_{\mathtt{Int}}^{5}, \cdots]
\end{equation*}
We denote the start and end instructions of a candidate sequence by $S$ and $E$, respectively.
In our example, the six element sequence starts with the first \texttt{push} instruction, and
terminates with the {pop} instruction terminates:
\begin{align*}
  S &:= \mathtt{push}_{\mathtt{Int}}^{0} \\
  E &:= \mathtt{pop}_{\mathtt{Int}}^{5}
\end{align*}

\paragraph{Unboxed Instruction Derivatives}

Analogous to the previous partial evaluation step we use to obtain the typed instructions
operating exclusively on boxed objects, we can use the same strategy to derive the even more
optimized derivatives.
For regular operations, such as our \texttt{intAdd} example, this is simple and straightforward, as
we just replace the boxed object representation \texttt{Value} by its native machine equivalent
\texttt{Int}:
\begin{lstlisting}[style=prettyhaskell]
intAdd' :: Int -> Int -> Int
intAdd' x y =  x + y
\end{lstlisting}
As a result, the compiler will generate an efficient native machine addition instruction and
completely sidestep the resolving of dynamic types, as well as (un-)boxing and reference counting
operations.

The problem with this \texttt{intAdd'} instruction derivative is, however, that we cannot perform a
type check on the operands anymore, as the implementation only allows operands of type
\texttt{Int}.
In consequence, for preserving semantics of the original interpreter instructions, we need a new
strategy for type checking.
Our solution to this problem is to bundle type checks and unboxing operations together via function
composition and move them to the load instructions which push unboxed operands onto the stack.
Assuming that we modify the declaration of \texttt{Stack} to contain heterogeneous data elements
(i.e., not only a list of \texttt{Value}, but also unboxed native machine words, denoted by
$\mathfrak{S}$), pushing an unboxed integer value onto the operand stack looks like this:
\begin{lstlisting}[style=prettyhaskell]
pushInt :: Value -> $\mathfrak{S}$ -> $\mathfrak{S}$
pushInt v s =
  let
    unboxConvert= $\mathfrak{c}_{\mathtt{int64}} \cdot \mathfrak{m}_{\mathtt{VInt}}$
  in
   case v of
     VInt value -> (unboxConvert value) : s
     _          -> -- misspeculation, generalize
\end{lstlisting}

Since the operand passing convention is type dependent, we cannot use the previous implementation of
\texttt{binaryOp} anymore, and need a typed version of it:
\begin{lstlisting}[style=prettyhaskell]
binaryIntOp :: (Int -> Int -> Int) -> $\mathfrak{S}$ -> $\mathfrak{S}$
binaryIntOp f s =
  let
    popInt = $\mathfrak{c}^{-1}_{\mathtt{int64}} \cdot \mathtt{pop}$
    pushInt = $\mathfrak{c}_{\mathtt{int64}} \cdot \mathtt{f}$
    (x, s') = popInt s
    (y, s'') = popInt s'
  in
   (pushInt x y) : s''
\end{lstlisting}

Finally, we need to make sure that once we leave an optimized sequence of instructions, the higher
level instruction sets continue to function properly.
Hence, we need to box all objects the sequence computes at the end of the optimized sequence.
For example, if we have a store instruction that saves a result into the environment, we need to
add boxing to its implementation:
\begin{lstlisting}[style=prettyhaskell]
storeIntOp :: $\mathfrak{S}$ -> Env -> String -> $\mathfrak{S}$
storeIntOp s e ident =
  let
    boxPopInt = $\mathfrak{m}^{-1}_{\mathtt{VInt}} \cdot \mathfrak{c}^{-1}_{\mathtt{int64}} \cdot \mathtt{pop}$
    (obj, s') = boxPopInt s
  in
    (\x (update e ident obj)) -> s'
\end{lstlisting}

\paragraph{Generalizing When Speculation Fails}
In our example of the \texttt{pushInt} interpreter instruction, we see on the last line that there
is a case when speculation fails.
A specific occurrence of \texttt{pushInt} verifies that the operand matches an expected type such
that subsequent instructions can modify on its unboxed representation.
Therefore, once the interpreter detects the misspeculation, we know that we have to invalidate all
subsequent instructions that speculate on that specific type.

The interpreter can back out of the misspeculation and resume sound interpretation by i) finding the
start of the speculatively optimized sequence, and ii) generalizing all specialized instructions up
by at least one level.
Both of these steps are trivial to implement, particularly since the instruction derivation
steps result in having three separate instruction sets.
Hence, the first step requires to identify the first instruction $i$ that does \emph{not} belong
to the current instruction set, which corresponds to the predecessor of the start instruction $S$
identified by our abstract interpreter.
The second step requires that we map each instruction starting at offset $i+1$ back to its more
general parent instruction---a mapping we can create and save when we create the initial mapping
from an instruction $I$ to its derivative $I'$.

To be sound, this procedure requires that we further restrict our abstract interpreter to identify
only sequences that have no side-effects.
As a result, we eliminate candidate sequences that have call instructions in between.
This is however only an implementation limitation and not an approach limitation, since all
non-local side-effects possible through function calls do not interfere with the current execution.
Therefore, we would need to ensure that we do not re-execute already executed functions and instead
push the boxed representation of a function call's result onto the operand stack.

\subsection{Putting It All Together}\label{ss:overview}

\newcommand{\nmBox}{$\beta$}
\newcommand{\nmUnbox}{$\beta^{-1}$}

\newcommand{\mappingFun}{$\mathfrak{m}$}
\newcommand{\invMappingFun}{$\mathfrak{m}^{-1}$}

\newcommand{\typedMappingFun}{$\mathfrak{m}_{\pi}$}
\newcommand{\typedInvMappingFun}{$\mathfrak{m}_{\pi}^{-1}$}

\newcommand{\pythonType}{T}
\newcommand{\nmType}{$\tau$}

\newcommand{\incaLabel}[1]{\textcolor{blue}{\texttt{#1}}}
\newcommand{\namasteLabel}[1]{\textcolor{SeaGreen}{\texttt{NAMA_FLOAT_}}\texttt{#1}}
\newcommand{\typeLabel}[1]{$\mathcal{#1}$}
\newcommand{\floatTypeLabel}{\typeLabel{F}}
\newcommand{\untypedLbl}[1]{\texttt{#1}}

\DeclareRobustCommand{\captionStepNr}[1]{\tikz[baseline=-.7ex,scale=.8, every node/.style={scale=.8}]\node[draw, shape=circle, fill=white] {#1};}

\tikzset{node distance=1.75cm, auto}
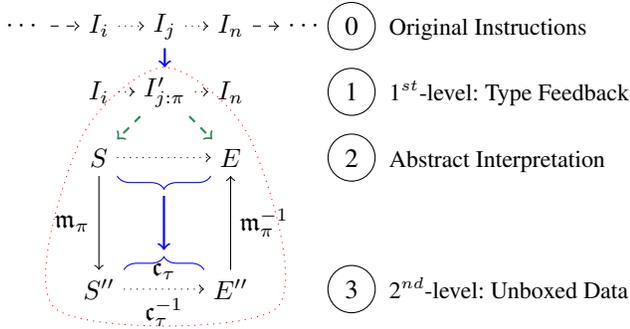
\begin{figure}[t!]
  \centering
  \begin{tikzpicture}
    \node (S) {$S$};
    \node (E) [right of=S] {$E$};
    \node (S1) [below of=S] {$S''$};
    \node (E1) [below of=E] {$E''$};
    \node (S0) [above of=S] {$I_{i}$};
    \node (E0) [above of=E] {$I_{n}$};
    \node (I)  at ($(S0)!0.5!(E0)$) {$I_{j}$};

    \node (S0a) at ($(S0)!0.5!(S)$) {$I_{i}$};
    \node (E0a) at ($(E0)!0.5!(E)$) {$I_{n}$};

    \node (Ideriv)  at ($(S0)!0.5!(E)$) {$I_{j:\pi}'$};

    \draw[dotted,red] plot[smooth cycle] coordinates{
      ($(Ideriv.north)$)
      ($(E.north east)$)
      ($(E1.south east)+(0.25,0)$)
      ($(S1.south west)+(-0.25,0)$)
      ($(S.north west)$)
    };

    \node (beforeS) [node distance=1cm, left of=S0] {$\cdots$};
    \node (afterE) [node distance=1cm, right of=E0] {$\cdots$};

    \node[draw, shape=circle, fill=white] (step0) at ($(afterE.east)+(.25cm,0)$) {0};
    \node[draw, shape=circle, fill=white] (step1) at ($(step0)-(0,0.875cm)$) {1};
    \node[draw, shape=circle, fill=white] (step2) at ($(step1)-(0,0.875)$) {2};
    \node[draw, shape=circle, fill=white] (step3) at ($(step2)-(0,1.75)$) {3};
    \node[anchor=text] (desc0) at ($(step0.base)+(.5cm,0)$) {{\small Original Instructions}};
    \node[anchor=text] (desc1) at ($(step1.base)+(.5cm,0)$) {{\small 1$^{st}$-level: Type Feedback}};
    \node[anchor=text] (desc2) at ($(step2.base)+(.5cm,0)$) {{\small Abstract Interpretation}};
    \node[anchor=text] (desc3) at ($(step3.base)+(.5cm,0)$) {{\small 2$^{nd}$-level: Unboxed Data}};

    \draw[->, dashed] (beforeS) to node {} (S0);
    \draw[->, dashed] (E0) to node {} (afterE);

    \draw[->, dotted] (S0)  to node {} (I);
    \draw[->, dotted] (I) to node {} (E0);

    \draw[->, dotted] (S0a)  to node {} (Ideriv);
    \draw[->, dotted] (Ideriv) to node {} (E0a);

    \draw[->, blue,thick] (I) to node {} (Ideriv);
    \draw[->, dashed, SeaGreen, thick] (Ideriv) to node {} (S);
    \draw[->, dashed, SeaGreen, thick] (Ideriv) to node {} (E);


    \draw[->] (S) to node  [swap] {\typedMappingFun{}} (S1);
    \draw[->] (E1) to node [swap] {\typedInvMappingFun{}} (E);

    \draw[->, dotted] (S)  to node {} (E);
    \draw[->, dotted] (S1) to node[above] {$\mathfrak{c}_{\tau}$} node[below] {$\mathfrak{c}_{\tau}^{-1}$} (E1);

    \draw[decorate,blue,decoration={brace,mirror,amplitude=5pt}] (S.south east)--(E.south west);
    \draw[decorate,blue,decoration={brace,amplitude=5pt}] (S1.north east)--(E1.north west);

    \draw[->, blue, thick] ($(S.south east)!0.5!(E.south west) + (0,-7.5pt)$)
          to ($(S1.north east)!0.5!(E1.north west)+(0,+7.5pt)$);

  \end{tikzpicture}
  \caption{Speculatively staged interpreter optimization using multi-level quickening.}
  \label{fig:mlq-overview}
\end{figure}

\autoref{fig:mlq-overview} shows, in a general form, how speculative staging of interpreter
optimizations works.
The first part of our technique requires speculatively staging optimized interpreter instructions at
interpreter compile-time.
In~\autoref{fig:mlq-overview}, the enclosing shaded shape highlights these staged instructions.
We systematically derive these optimized interpreter instructions from the original interpreter
instructions.
The second part of our technique requires run-time information and concerts the speculatively staged
interpreter instructions such that optimized interpretation preserves semantics.

The interpreter starts out executing instructions belonging to its original instruction set
(see~\captionStepNr{0} in~\autoref{fig:mlq-overview}).
During execution, the interpreter can capture type feedback by rewriting itself to optimized
instruction derivatives:
\autoref{fig:mlq-overview} shows this in step~\captionStepNr{1}, where instruction $I'_{j:\pi}$
replaces the more generic instruction $I_{j}$, thereby capturing the type information $\pi$ at
offset $j$.
The next step,~\captionStepNr{2},  propagates the captured type information $\pi$ to a complete
sequence of instructions, starting with instruction $S$ and terminating in instruction $E$.
We use an abstract interpreter to identify candidate sequences operating on the same type.
Once identified,~\captionStepNr{3} of~\autoref{fig:mlq-overview} illustrates how we rewrite a
complete instruction sequence $S,\ldots, E$ to optimized instruction derivatives $S'',\dots,E''$.
This third instruction set operates on unboxed native machine data types, and therefore requires
generic ways to handle (un-)boxing of objects of type $\pi$ (cf.~$\mathfrak{m}_{\pi}$ and
$\mathfrak{m}_{\pi}^{-1}$), as well as converting data to and from the operand stack
(cf.~$\mathfrak{c}_{\tau}$ and $\mathfrak{c}_{\tau}^{-1}$).
All instructions circled by the dotted line of~\autoref{fig:mlq-overview} are instruction
derivatives that we speculatively stage at interpreter compile-time.

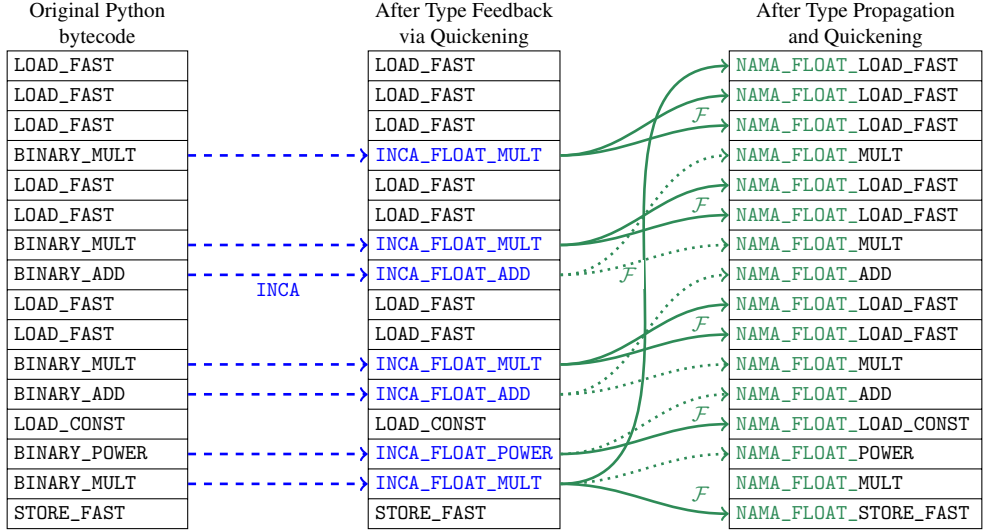
\begin{figure*}[t!]
  \centering
  \begin{tikzpicture}[scale=.8, every node/.style={scale=.8}]
    \begin{scope}
      \node [draw, align=right, anchor=text, rectangle split, minimum width=3cm,
             rectangle split  parts=16, rectangle split part align={left}]
      (Orig) at (0,0) {
        \texttt{LOAD_FAST}
        \nodepart{two}
        \texttt{LOAD_FAST}
        \nodepart{three}
        \texttt{LOAD_FAST}
        \nodepart{four}
        \texttt{BINARY_MULT}
        \nodepart{five}
        \texttt{LOAD_FAST}
        \nodepart{six}
        \texttt{LOAD_FAST}
        \nodepart{seven}
        \texttt{BINARY_MULT}
        \nodepart{eight}
        \texttt{BINARY_ADD}
        \nodepart{nine}
        \texttt{LOAD_FAST}
        \nodepart{ten}
        \texttt{LOAD_FAST}
        \nodepart{eleven}
        \texttt{BINARY_MULT}
        \nodepart{twelve}
        \texttt{BINARY_ADD}
        \nodepart{thirteen}
        \texttt{LOAD_CONST}
        \nodepart{fourteen}
        \texttt{BINARY_POWER}
        \nodepart{fifteen}
        \texttt{BINARY_MULT}
        \nodepart{sixteen}
        \texttt{STORE_FAST}
      };
    \end{scope}

    \begin{scope}
      \node [draw, align=right, anchor=text, rectangle split, minimum width=3cm,
             rectangle split  parts=16, rectangle split part align={left}]
      (Inca) at (6,0) {
        \texttt{LOAD_FAST}
        \nodepart{two}
        \texttt{LOAD_FAST}
        \nodepart{three}
        \texttt{LOAD_FAST}
        \nodepart{four}
        \incaLabel{INCA_FLOAT_MULT}
        \nodepart{five}
        \texttt{LOAD_FAST}
        \nodepart{six}
        \texttt{LOAD_FAST}
        \nodepart{seven}
        \incaLabel{INCA_FLOAT_MULT}
        \nodepart{eight}
        \incaLabel{INCA_FLOAT_ADD}
        \nodepart{nine}
        \texttt{LOAD_FAST}
        \nodepart{ten}
        \texttt{LOAD_FAST}
        \nodepart{eleven}
        \incaLabel{INCA_FLOAT_MULT}
        \nodepart{twelve}
        \incaLabel{INCA_FLOAT_ADD}
        \nodepart{thirteen}
        \texttt{LOAD_CONST}
        \nodepart{fourteen}
        \incaLabel{INCA_FLOAT_POWER}
        \nodepart{fifteen}
        \incaLabel{INCA_FLOAT_MULT}
        \nodepart{sixteen}
        \texttt{STORE_FAST}
      };
    \end{scope}

    \begin{scope}
      \node [draw, align=right, anchor=text, rectangle split, minimum width=3cm,
             rectangle split  parts=16, rectangle split part align={left}]
      (Namaste) at (12,0) {
        \namasteLabel{LOAD_FAST}
        \nodepart{two}
        \namasteLabel{LOAD_FAST}
        \nodepart{three}
        \namasteLabel{LOAD_FAST}
        \nodepart{four}
        \namasteLabel{MULT}
        \nodepart{five}
        \namasteLabel{LOAD_FAST}
        \nodepart{six}
        \namasteLabel{LOAD_FAST}
        \nodepart{seven}
        \namasteLabel{MULT}
        \nodepart{eight}
        \namasteLabel{ADD}
        \nodepart{nine}
        \namasteLabel{LOAD_FAST}
        \nodepart{ten}
        \namasteLabel{LOAD_FAST}
        \nodepart{eleven}
        \namasteLabel{MULT}
        \nodepart{twelve}
        \namasteLabel{ADD}
        \nodepart{thirteen}
        \namasteLabel{LOAD_CONST}
        \nodepart{fourteen}
        \namasteLabel{POWER}
        \nodepart{fifteen}
        \namasteLabel{MULT}
        \nodepart{sixteen}
        \namasteLabel{STORE_FAST}
      };
    \end{scope}

    \node[align=center] at ($(Orig.north)   +(0, 12pt)$) (origLabel) {Original Python\\bytecode};
    \node[align=center] at ($(Inca.north)   +(0, 12pt)$) (incaLabel) {After Type Feedback\\via Quickening};
    \node[align=center] at ($(Namaste.north)+(0, 12pt)$) (namaLabel) {After Type Propagation\\and Quickening};



    \newcommand{\incaQuickening}[2]{ \draw[->, blue, dashed, line width=1pt] (Orig.#1 east) to node [below] {#2} (Inca.#1 west) }
    \incaQuickening{four}{};
    \incaQuickening{seven}{};
    \incaQuickening{eight}{\inca{}};
    \incaQuickening{eleven}{};
    \incaQuickening{twelve}{};
    \incaQuickening{fourteen}{};
    \incaQuickening{fifteen}{};

    \newcommand{\nmTypeProp}[2]{        \draw[->, SeaGreen, line width=1pt] (Inca.#1 east) to [out=0, in=180] (Namaste.#2 west) }
    \newcommand{\nmLabeledTypeProp}[2]{ \draw[->, SeaGreen, line width=1pt] (Inca.#1 east) to [out=0, in=180] node [above right, near end] {\floatTypeLabel{}} (Namaste.#2 west) }
    \newcommand{\nmTrivial} [1]{        \draw[->, SeaGreen, dotted, line width=1pt] (Inca.#1 east) to (Namaste.#1 west) }
    \newcommand{\nmResults} [2]{        \draw[->, SeaGreen, dotted, line width=1pt] (Inca.#1 east) to [out=0, in=180] (Namaste.#2 west) }

    \nmTypeProp{four}{two};
    \nmLabeledTypeProp{four}{three};

    \nmTypeProp{seven}{five};
    \nmLabeledTypeProp{seven}{six};

    \nmTypeProp{eleven}{nine};
    \nmLabeledTypeProp{eleven}{ten};

    \nmLabeledTypeProp{fourteen}{thirteen};
    \draw[->, SeaGreen, line width=1pt] (Inca.fifteen  east)     to [out=0, in=180] node [midway,fill=white] {\floatTypeLabel{}}  (Namaste.text west);
    \nmLabeledTypeProp{fifteen}{sixteen};

    \nmResults{eight}{seven};
    \nmResults{eight}{four};

    \nmResults{twelve}{eleven};
    \nmResults{twelve}{eight};

    \nmResults{fourteen}{twelve};
    \nmResults{fifteen}{fourteen};

  \end{tikzpicture}
  \caption{Concrete Python bytecode example and its step-wise optimization using multi-level
    quickening for concerting at run-time.}
  \label{fig:orig-inca-namaste}
\end{figure*}

\section{Implementation}\label{s:implementation}

This section presents implementation details of how we use speculative staging of optimized
interpreter instructions to substantially optimize execution of a Python 3 series interpreter.
This interpreter is an implementation vehicle that we use to demonstrate concrete instantiations of
our general optimization framework.
We use an example sequence of Python instructions to illustrate both the abstract interpretation as
well as deriving the optimized interpreter instructions.
Python itself is implemented in C and we use casts to force the compiler to use specific semantics.

\autoref{fig:orig-inca-namaste} shows our example Python instruction sequence and how we
incrementally and iteratively rewrite this sequence using our speculatively staged optimized
interpreter instruction derivatives.
We use the prefix \texttt{INCA} to indicate optimized interpreter instruction derivatives used for
inline caching, i.e., the first-level quickening.
The third instruction set uses the \texttt{NAMA} prefix, which abbreviates native machine execution
since all instructions directly operate on machine data and hence use efficient machine instructions
to implement their operation semantics.
This corresponds to the second-level quickening.
Note that the~\texttt{NAMA} instruction set is portable by construction, as it leverages the
back-end of the ahead-of-time compiler at interpreter compile-time.

\begin{figure}[t!]
  \centering
  \begin{tikzpicture}[grow=up, level/.style={level distance=1cm, sibling distance=3cm}, scale=.6, every node/.style={scale=.6}]]
    \tikzstyle{aitree}=[draw, rectangle]
    \tikzstyle{arrUp}=[edge from parent/.style={draw,-latex}]
    \tikzstyle{arrDown}=[edge from parent/.style={draw,latex-}]
    \tikzstyle{level 2}=[sibling distance=6cm]
    \tikzstyle{level 4}=[sibling distance=6cm]

    \node[aitree] (storeFast) {\untypedLbl{STORE_FAST}}
    child[arrDown] {
      node[aitree] (mult0) {\incaLabel{INCA_FLOAT_MULT}}
      child {
        node[aitree] (power) {\incaLabel{INCA_FLOAT_POWER}}
        child[arrUp] { node[aitree] (const) {\untypedLbl{LOAD_CONST}} }
        child {
          node[aitree] (add0) {\incaLabel{INCA_FLOAT_ADD}}
          child {
            node[aitree] (mult1) {\incaLabel{INCA_FLOAT_MULT}}
            child[arrUp] { node[aitree] (load1) {\untypedLbl{LOAD_FAST}} }
            child[arrUp] { node[aitree] (load2) {\untypedLbl{LOAD_FAST}} }
          }
          child[] {
            node[aitree] (add2) {\incaLabel{INCA_FLOAT_ADD}}
            child {
              node[aitree] (mult3) {\incaLabel{INCA_FLOAT_MULT}}
              child[arrUp] { node[aitree] (load3) {\untypedLbl{LOAD_FAST}} }
              child[arrUp] { node[aitree] (load4) {\untypedLbl{LOAD_FAST}} }
            }
            child[sibling distance=7cm] {
              node[aitree] (mult4) {\incaLabel{INCA_FLOAT_MULT}}
              child[arrUp] { node[aitree] (load5) {\untypedLbl{LOAD_FAST}} }
              child[arrUp] { node[aitree] (load6) {\untypedLbl{LOAD_FAST}} }
            }
          }
        }
      }
      child[arrUp] { node[aitree] (load0) {\untypedLbl{LOAD_FAST}} }
    };

    \tikzstyle{pattern}=[draw, circle, above, solid, thin]
  \end{tikzpicture}
  \caption{Data-flow of the sequence of instructions of~\autoref{fig:orig-inca-namaste} in abstract
    interpretation. Arrows indicate direction of type propagation.}
  \label{fig:ai-tree}
\end{figure}
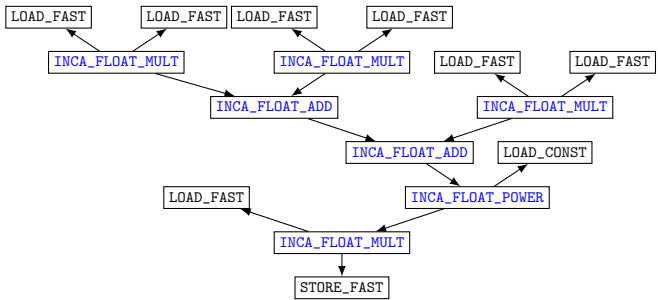

The \loadFast{} instruction pushes a local variable onto the operand stack, and conversely
\storeFast{} pops an object off the operand stack and stores it in the local stack frame.
\autoref{tab:start-end-instrs} lists the set of eligible start and end instructions for our abstract
interpreter, and~\autoref{fig:ai-tree} illustrates the data flow of the instruction sequence as
assembled by the abstract interpreter.
In our example, the whole instruction sequence operates on a single data type, but in general the
abstract interpreter needs to be aware of the type lattice implemented by the Python interpreter.
For example, dividing two long numbers results in a float number and comparing two complex numbers
results in a long number.
We model these type conversions as special rules in our abstract interpreter.

\begin{table}[t!]
  \centering
  \begin{tabular}{|l|l|}
    \hline
    Start Instr. & End Instr. \\
    \hline
    \hline
    \coldata{LOAD_CONST} & \coldata{POP_JUMP_IF_FALSE} \\
    \coldata{LOAD_FAST}  & \coldata{POP_JUMP_IF_TRUE}  \\
    & \coldata{RETURN_VALUE}      \\
    & \coldata{STORE_FAST}        \\
    & \coldata{YIELD_VALUE}       \\
    \hline
  \end{tabular}
  \caption{Valid start and end instructions used for abstract interpretation.\label{tab:start-end-instrs}}
\end{table}

Having identified a complete sequence of instructions that all operate on operands of type
\pyType{Float}, we can replace all instructions of this sequence with optimized derivatives directly
operating on native machine \texttt{float} numbers.

\subsection{Deriving Optimized Python Interpreter Instructions}\label{ss:quickening}

In this section we present details to implement speculative staging of optimized interpreter
instructions in a methodological fashion.
First, we are going describe the functions we use to (un-)box Python objects and the conventions we
use to pass operand data on the mixed-value operand stack.
We treat arithmetic operations for Python's integers, floating point and complex numbers, though our
technique is not limited to these.
Second, we illustrate the derivation steps in concrete Python code examples---while in theory we
could use a partial evaluator to generate the derivatives, the manual implementation effort is so
low that it does not justify using a partial evaluator.

\subsubsection{(Un-)boxing Python Objects}

The Python interpreter supports uniform procedures to box and unbox Python objects to native machine
numbers, and we just briefly mention which functions to use and give their type.

\paragraph{Integers}
Python represents its unbounded range integers by the C \texttt{struct} \pyType{Long}.
\begin{align*}
  \mathfrak{m}_{\mathtt{PyLong\_Type}} & := \mathtt{PyLong\_AS\_LONG} : \mathtt{PyLongObject*} \rightarrow \mathtt{int64} \\
  \mathfrak{m}_{\mathtt{PyLong\_Type}}^{-1} & := \mathtt{PyLong\_FROM\_LONG} :  \mathtt{int64} \rightarrow \mathtt{PyLongObject*}
\end{align*}

\paragraph{Floating Point Numbers}
Floating point numbers are represented by the C \texttt{struct} \pyType{Float}.
\begin{align*}
  \mathfrak{m}_{\mathtt{PyFloat\_Type}} & := \mathtt{PyFloat\_AS\_DOUBLE} : \mathtt{PyFloatObject*} \rightarrow \mathtt{double} \\
  \mathfrak{m}_{\mathtt{PyFloat\_Type}}^{-1} & := \mathtt{PyFloat\_FromDouble} :  \mathtt{double} \rightarrow \mathtt{PyFloatObject*}
\end{align*}

\paragraph{Complex Numbers}
Complex numbers are represented by the C \texttt{struct} \pyType{Complex}, but we cannot directly
access a native machine representation of its data.
This is due to a complex number consisting of two parts, a real and an imaginary part:
\begin{lstlisting}[language=C,style=othercode]
typedef struct {
    double real;
    double imag;
} complex_t;
\end{lstlisting}
Furthermore, we need to know something about the internals of the \pyType{Complex} implementation to
access the native machine data.
Internally, Python uses a similar struct to \texttt{complex_t} (\texttt{Py_complex}) to hold the
separate parts and we can access the data via \texttt{((PyComplexObject*)x)->cval}.
\begin{align*}
  \mathfrak{m}_{\mathtt{PyComplex\_Type}} & := \mathtt{PyComplexObject*} \rightarrow (\mathtt{double}, \mathtt{double}) \\
  \mathfrak{m}_{\mathtt{PyComplex\_Type}}^{-1} & := (\mathtt{double}, \mathtt{double}) \rightarrow \mathtt{PyComplexObject*}
\end{align*}

\subsubsection{Operand Stack Data Passing Convention}

Across all three instruction sets, instructions pass operands using the operand stack.
The original Python instruction set operates exclusively on Python objects which reside on the heap,
i.e., the operand stack passes pointers to the heap, of type \texttt{uint64} on a 64-bit machine.
The lowest level instruction set operates on native machine data, i.e., we need to map native
machine data types to the existing operand stack.
We rely on C constructs of explicit casts and \texttt{union}s to ensure that we attach the proper
semantics to the bits passed around.

\paragraph{Passing Integers}
Naturally, it is trivial in C to support passing integers on the mixed operand stack: we simply cast
from the signed integer representation to the unsigned representation used for pointers, too.
\begin{align*}
  \mathfrak{c}_{\mathtt{int64}}(\mathtt{o}: \mathtt{int64})      & := \mathtt{(uint64)o} \\
  \mathfrak{c}_{\mathtt{int64}}^{-1}(\mathtt{o}: \mathtt{uint64}) & := \mathtt{(int64)o}
\end{align*}

\paragraph{Passing Floats}
To pass floating point numbers, we need to avoid implicit casting a C compiler would insert when
using a cast from \texttt{double} to \texttt{uint64}.
A solution to circumvent this, is to use a C \texttt{union} that allows to change the semantics the
compiler attaches to a set of bits.
More precisely, we use the following \texttt{union}:
\begin{lstlisting}[language=C, morekeywords={uint64,int64},style=othercode]
typedef union { uint64 word; double dbl; } map_t;
\end{lstlisting}
\texttt{Map_t} allows us to change the semantics by using the corresponding field identifier and
thus suffices to map \texttt{double}s to \texttt{uint64} representation in a transparent and
portable fashion.
\begin{align*}
  \mathfrak{c}_{\mathtt{double}}(\mathtt{o}: \mathtt{double})     & := (\mathtt{map\_t~m};\mathtt{m.dbl}= \mathtt{o};(\mathtt{uint64})\mathtt{m.word}) \\
  \mathfrak{c}_{\mathtt{double}}^{-1}(\mathtt{o}: \mathtt{uint64}) & := (\mathtt{map\_t~m};\mathtt{m.word}= \mathtt{o}; \mathtt{m.dbl})
\end{align*}

\paragraph{Passing Complex Numbers}
As previously described, we represent a complex number using two \texttt{double} numbers.
Therefore, we can reuse the functions that map floating point numbers:
\begin{align*}
  \mathfrak{c}_{\mathtt{complex}}&(\mathtt{o}: \mathtt{complex\_t})     := (\mathfrak{c}_{\mathtt{double}}(\mathtt{o.real}), \mathfrak{c}_{\mathtt{double}}(\mathtt{o.imag}))\\
  \mathfrak{c}_{\mathtt{complex}}^{-1}&(\mathtt{i}: \mathtt{uint64}, \mathtt{r}: \mathtt{uint64}) :=
  (\mathtt{complex\_t~c}; \\
  & \mathtt{c.real}= \mathfrak{c}_{\mathtt{double}}^{-1}(\mathtt{r}); \mathtt{c.imag}= \mathfrak{c}_{\mathtt{double}}^{-1}(\mathtt{i}))
\end{align*}
But, the operand stack passing convention alone does not suffice since passing native machine
complex numbers requires two stack slots instead of one.
Consequently, we double the operand stack size such that all instruction operands on the stack could
be two-part complex numbers.
Since the abstract interpreter identifies whole sequences of instructions, there is always a
termination instruction that boxes the two-part double numbers into a \texttt{PyComplexObject}
instance.
As a result, this temporary use of two operand stack slots is completely transparent to  all
predecessor instructions of the sequence as well as all successor instructions of the sequence.

\subsubsection{Example Instructions}\label{ss:examples}

The previous section contains all details necessary to inductively construct all optimized
interpreter instructions that modify native machine data types.
In this section, we give concrete Python interpreter instruction implementation examples for
completeness.

\paragraph{Original Python Instruction}

The following program excerpt shows Python's original implementation of the arithmetic subtraction
operation:
\begin{lstlisting}[language=C,style=othercode]
case BINARY_SUBTRACT:
    w = POP();
    v = TOP();
  BINARY_SUBTRACT_MISS:
    x = PyNumber_Subtract(v, w);
    Py_DECREF(v);
    Py_DECREF(w);
    SET_TOP(x);
    if (x != NULL) DISPATCH();
    goto on_error;
\end{lstlisting}
The resolving procedure of dynamic types is not visible and resides in the
\texttt{PyNumber_Subtract} function.
The resolving is much more complicated than our simplified interpreter substrate suggests, in
particular due to the presence of ad-hoc polymorphism and inheritance.
In our simplified interpreter substrate all dynamic types were known at compile-time and we could
use pattern matching to express semantics properly and exhaustively.
For languages such as Python, this is in general not possible.
For example, one could use operator overloading for integers to perform subtraction, or perform the
traditional integer addition by inheriting from the system's integers.

\paragraph{First-level Quickening}

By fixing operands \texttt{v} and \texttt{w} to the type \pyType{Float}, we can derive the following
optimized instruction derivative, expressed as \texttt{INCA_FLOAT_SUBTRACT}.
\begin{lstlisting}[language=C,mathescape=true,style=othercode]
case INCA_FLOAT_SUBTRACT:
    w= POP();
    v= TOP();
    if (!$\textcolor{red}{\mathfrak{T}}$(v, w, PyFloat_Type)) {
        /* misspeculation, generalize */
        goto BINARY_SUBTRACT_MISS;
    } // if
    x= PyFloat_Type.tp_as_number->nb_sub(v, w);
    Py_DECREF( w );
    Py_DECREF( v );
    SET_TOP(x);
    if (x != NULL) DISPATCH();
    goto on_error;
\end{lstlisting}
On lines four to seven, we see what happens on misspeculation.
After the type check (stylized by symbol $\textcolor{red}{\mathfrak{T}}$) fails, we resume execution
of the general instruction that \texttt{INCA_FLOAT_SUBTRACT} derives from, \texttt{BINARY_SUBTRACT}
in this case.
We change the implementation of \texttt{BINARY_SUBTRACT} to add another label that we can use for
resuming correct execution.

Directly calling \texttt{nb_sub} on \pyType{Float} optimizes the complex resolving of dynamic typing
hinted at before.

Furthermore, this type-specialized instruction derivative illustrates that we can in fact infer that
both operands as well as the result (\texttt{x}) are of type \pyType{Float}.

\paragraph{Second-level Quickening}

The second-level quickening step optimizes sequences of interpreter instructions to directly modify
native machine data.
Hence, we move the required type check to the corresponding load instruction:
\begin{lstlisting}[mathescape=true,language=C,morekeywords={uint64,int64,TARGET,PUSH,NEXT_INSTR,T},style=othercode]
case NAMA_FLOAT_LOAD_FAST:
    PyObject *x= fastlocals[oparg];
    map_t result;
    if ($\textcolor{red}{\mathfrak{T}}$(x, PyFloat_Type))
        result.dbl= PyFloat_AS_DOUBLE(x);
    else /* misspeculation, generalize */
    PUSH( result.word );
    NEXT_INSTR();
\end{lstlisting}
The corresponding floating point subtract operation need not perform any type checks, reference
counting, or (un-)boxing operations anymore:
\begin{lstlisting}[language=C,mathescape=true,style=othercode,morekeywords={uint64,map_t}]
case NAMA_FLOAT_SUBTRACT:
    w= POP();
    v= TOP();
    {
        map_t s, t;
        s.word= (uint64)v;
        t.word= (uint64)w;
        s.dbl-= t.dbl;
        SET_TOP( s.word );
        DISPATCH();
    }
\end{lstlisting}

Both of the native floating point interpreter instructions make use of the \texttt{map_t}
union as previously explained to avoid implicit conversions as emitted by the compiler.

In general, the type-specific load instructions have to validate all assumptions needed to
manipulate unboxed native machine data.
For example, integers in Python 3 are by default unbounded, i.e., they can exceed native machine
bounds.
As a result, we modify the corresponding integer load instruction to check whether it is still safe
to unbox the integer.
However, it is worth noting that these are only implementation limitations, as for example we could
expand this technique to use two machine words and perform a 128-bit addition because we already
doubled the stack size to accommodate complex numbers.

\paragraph{Implementation Remarks}
The first-level quickening step has already been explored in 2010 by
Brunthaler~\cite{brunthaler+10a,brunthaler+10b}, and we used his publicly available
implementation~\cite{brunthaler+12} as a basis for ours.
In the remainder of this paper, we refer to the original interpreter as \inca{} (an abbreviation for
inline caching), and the modified interpreter that supports the new optimizations as \mlq{}, which
is short for multi-level quickening.

We use a simple code generator written in Python that generates the C code for \emph{all}
instructions of the Python interpreter.
We run our code generator as a pre-compile step when compiling the interpreter, and rely on the
existing build infrastructure to build the interpreter.
In consequence, all of the instruction derivatives speculatively added to the interpreter and
available for concerting at run-time; sidestepping dynamic code generation altogether.

We provide templates of the C instructions using the language of the Mako template
engine~\cite{bayer+13}.
The semantics of all instruction derivatives is essentially identical, e.g., adding numbers, which
is why derivative instructions are merely optimized copies operating on specialized structures or
types.
Hence, these templates capture all essential details and help keeping redundancy at bay.
If we were to create a domain-specific language for generating interpreters, similar to the
VMgen/Tiger~\cite{ertl.etal+02,casey.etal+05a} project, we could express the derivative relationship
for instructions, thereby reducing the costs for creating these templates.
The next section provides lines-of-code data regarding our interpreter implementation generator
(see~\autoref{ss:interpreter-data}).

\section{Evaluation}\label{s:evaluation}

\paragraph{Systems and Procedure}
We ran the benchmarks on an Intel Nehalem i7-920 based system running at a frequency of 2.67 GHz, on
Linux kernel version 3.0.0-26 and \texttt{gcc} version 4.6.1.
To minimize perturbations by third party systems, we take the following precautions.
First, we disable Intel's Turbo Boost~\cite{intel+13a} feature to avoid frequency scaling based on
unpublished, proprietary heuristics.
Second, we use \texttt{nice -n -20} to minimize operating system scheduler effects.
Third, we use 30 repetitions for each pairing of a benchmark with an interpreter to get stable
results; we report the geometric mean of these repetitions, thereby minimizing bias towards
outliers.

\paragraph{Benchmarks}
We use several benchmarks from the computer language benchmarks game~\cite{fulgham+13}.
This is due to the following reasons.
First, since we are using a Python 3 series interpreter, we cannot use programs written for Python 2
to measure performance.
Unfortunately, many popular third party libraries and frameworks, such as Django, twisted,
etc. have not released versions of their software officially supporting Python 3.
Compatibility concerns aside, the Python community has no commonly agreed upon comprehensive
set of benchmarks identified to assess Python performance.
Second, some popular libraries have custom C code modules that perform computations.
Effectively, benchmarking these programs corresponds to measuring time \emph{not} spent in the
interpreter, and therefore would skew the results in the wrong direction.
Instead, we use the following benchmarks that stress raw interpreter performance:
\textsf{binarytrees},
\textsf{fannkuch},
\textsf{fasta},
\textsf{mandelbrot},
\textsf{nbody}, and
\textsf{spectralnorm}.

Finally, we rely on those benchmarks because they allow comparison with other implementations, such
as PyPy.
PyPy officially only supports Python 2, but since none of those benchmarks use Python 3
specifics---with the notable exception of \textsf{fannkuch}, which required minor changes---we run
the identical programs under PyPy.
This may sound like a contradiction, but is in fact only possible with the chosen set of programs
and cannot in general be expected to hold for other programs.

\subsection{Benchmark Results}\label{ss:benchmark-results}

\begin{figure*}[t!]
  \centering
  \includegraphics[width=.9\textwidth]{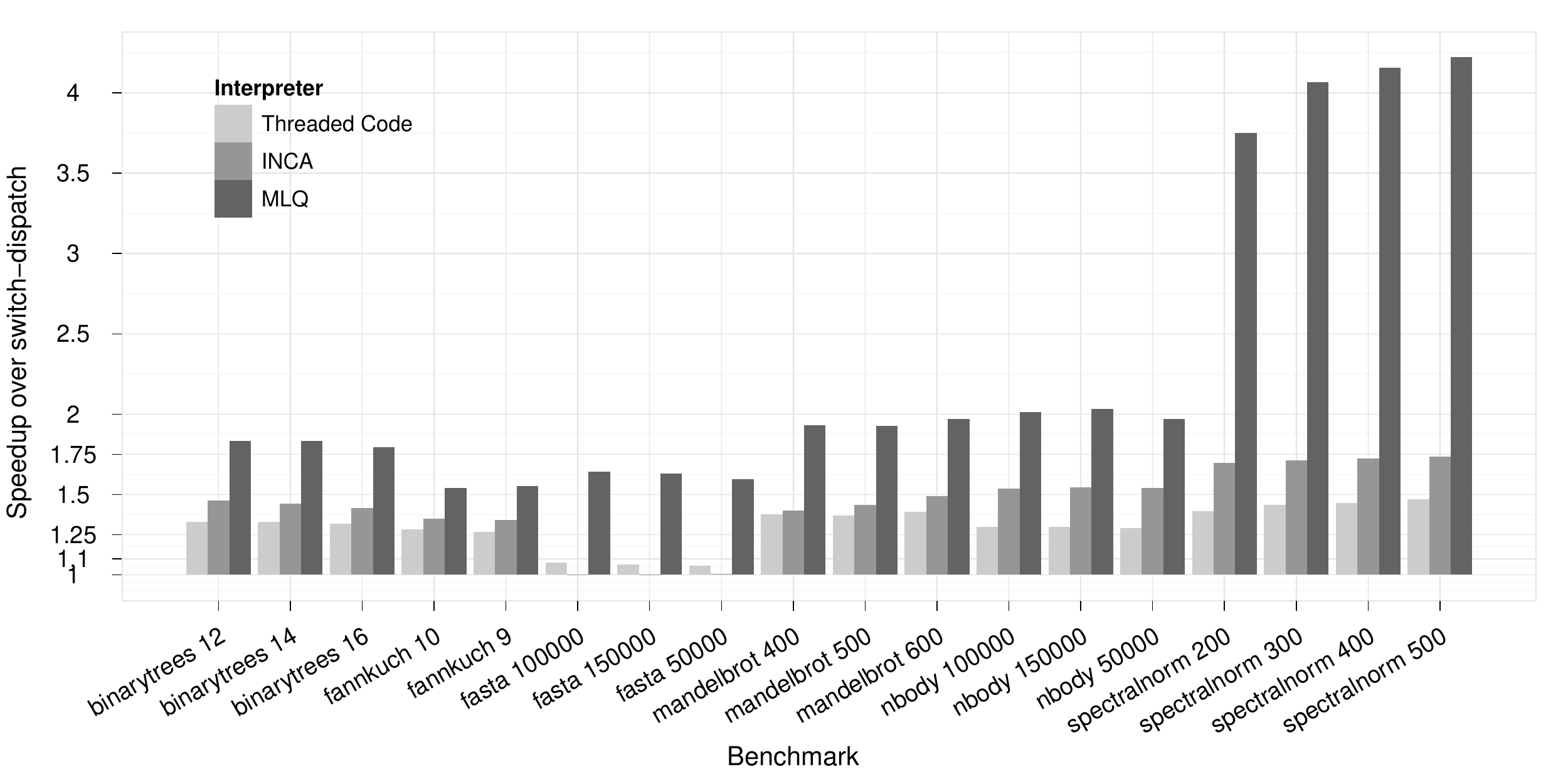}
  \caption{Detailed speedups per benchmark normalized by the CPython 3.2.3 interpreter using
    switch-dispatch.}
  \label{fig:benchmarks}
\end{figure*}

\autoref{fig:benchmarks} shows the performance results obtained on our test system.
We report a maximum speedup by up to a factor of \namasteMaxSU{} over the CPython 3.2.3 interpreter
using switch dispatch.
\inca{} itself improves performance by up to a factor of \incaMaxSU{}.
As a result, the \mlq{} system improves upon the previous maximum speedups by \fasterThanINCA{}.

\paragraph{PyPy Comparison}
Even though our speculatively staged \mlq{} interpreter is no match in comparison to the multi-year,
multi-person effort that went into the PyPy implementation (see implementation measurements in the
discussion in~\autoref{ss:discussion}), we want to give a realistic perspective of the potential of
\mlq{} interpretation.
Therefore, we evaluated the performance against PyPy version 1.9~\cite{pypy+13}.
Note that the times we measured include start-up and warm-up times for PyPy; since it is not clear
at which point in time the benefits of JIT compilation are visible.

\begin{table}[t!]
  \centering
  \begin{tabular}{|l|r|r|r|}
    \hline
    \multicolumn{1}{|c|}{Benchmark} & \multicolumn{1}{c|}{PyPy 1.9} & \multicolumn{1}{c|}{\mlq{}} & \multicolumn{1}{c|}{$\frac{\mathrm{PyPy}}{\mlq{}}$} \\
    \hline
    \hline
    \textsf{binarytrees}            & 3.2031 $\times$               & 1.8334 $\times$                   & 1.7471 \\
    \hline
    \textsf{fannkuch}               & 8.2245 $\times$               & 1.5240 $\times$                   & 5.2884 \\
    \hline
    \textsf{fasta}                  & 13.4537 $\times$              & 1.6906 $\times$                   & 7.9579 \\
    \hline
    \textsf{mandelbrot}             & 6.3224 $\times$               & 1.9699 $\times$                   & 3.2095 \\
    \hline
    \textsf{nbody}                  & 12.3592 $\times$              & 2.0639 $\times$                   & 5.9883 \\
    \hline
    \textsf{spectralnorm}           & 3.5563 $\times$               & 4.0412 $\times$                   & 0.8800 \\
    \hline
  \end{tabular}
  \caption{Speedups of PyPy 1.9 and our \mlq{} system normalized by the CPython 3.2.3 interpreter using switch-dispatch.\label{tab:pypy-speedups}}
\end{table}

\autoref{tab:pypy-speedups} lists the geometric mean of speedups per benchmark that we measured on
our Intel Nehalem system.
During our experiment we also measured overall memory consumption and report that our system uses
considerably less memory at run-time: PyPy uses about 20 MB, whereas our \mlq{} interpreter uses
less than 7 MB.
This is primarily due to the systems using different memory management techniques: \mlq{} uses
CPython's standard reference counting, whereas PyPy offers several state-of-the-art garbage
collectors, such as a semi-space copying collector and a generational garbage collector.
Surprisingly, we find that using a more powerful memory management technique does not automatically
translate to higher performance.
Since \mlq{} is particularly effective at eliminating the overhead of reference counting---reducing
required reference count operations, as well as using native-machine data instead of boxed
objects---we can take full advantage of its benefits: determinism and
space-efficiency~\cite{deutsch.bobrow+76}.

\subsection{Interpreter Data}\label{ss:interpreter-data}


We rely on David Wheeler's \texttt{sloccount} utility~\cite{wheeler+10} to measure lines of code.
For calculating the number of interpreter instructions, we use a regular expression to select the
beginning of the instructions and \texttt{wc -l} to count their occurrences.

\paragraph{Instruction-Set Extension} The CPython 3.2.3 interpreter has 100 interpreter
instructions spanning 1283 lines of code.
The \inca{} instruction set of the \inca{} interpreter adds another 53 instructions to the
interpreter, totaling 3050 lines of code (i.e., a plus of 138\% or 1767 lines of code).
\namaste{} itself requires additional 134 interpreter instructions adding another 1240 lines of code
(increase by 41\% over \inca{}).
We adapted the existing Python code generator of the \inca{} system to generate the \namaste{}
instruction derivatives' C implementation.
The original code generator has 2225 lines of Python code, where 1700 lines of code just reflect
the type structure code extracted from the C \texttt{struct}s of Python objects via \texttt{gdb}.
The required changes were about 600 lines of Python code, resulting in the updated code generator
having 2646 lines of Python code.
The \inca{} code generator uses 2225 lines of C code templates.
To support the \namaste{} instruction set, we added another 1255 lines of templatized C
code---giving 3480 lines of template code in total.
In addition to this, our abstract interpreter identifying eligible sequences requires around 400
lines of C code.

\paragraph{Portability}
As we have briefly mentioned before, our speculative staging leverages the existing backend of the
ahead-of-time compiler that is used to compile the interpreter.
Therefore, our technique is portable by construction, i.e., since we implemented our optimized
derivatives in C, the interpreter is as portable as any other C program.
We confirm this by compiling the optimized interpreter on a PowerPC system.
This did not require changing a single line of code.

\begin{table}[t!]
  \centering
    \begin{tabular}{|l|r|r|r|}
      \hline
      \multicolumn{1}{|c|}{Interpreter} & Binary Size & \multicolumn{2}{c|}{Increase} \\
      \cline{3-4}
      & (bytes)     & (kB) & ($\%$)                 \\
      \hline
      \hline
      Python 3.2.3/switch-dispatch      & 2,135,412   & 0    & 0.0                    \\
      Python 3.2.3/threaded code        & 2,147,745   & 12   & 0.6                    \\
      \mlq{} Python interpreter         & 2,259,616   & 121  & 5.8                    \\
      \hline
    \end{tabular}
    \caption{\mlq{} binary size increase without debug information on Intel Nehalem i7-920.\label{tab:binary-size-increase}}
\end{table}

\paragraph{Space Requirements}\autoref{tab:binary-size-increase} presents the effect of implementing
our speculatively staged \mlq{} Python interpreter on the binary size of the executable.
We see that going from a switch-based interpreter to a threaded code interpreter requires additional
12 kB of space.
Finally, we see that adding two additional instruction sets to our Python interpreter requires less
than 110 kB of additional space (when discounting the space requirement from threaded code).

\subsection{Discussion}\label{ss:discussion}

The most obvious take-away from~\autoref{fig:benchmarks} is that there is clearly a varying
optimization potential when using our optimization.
Upon close investigation, we found that this is due to our minimal set of eligible start and
end instructions (see~\autoref{tab:start-end-instrs}).
For example, there are other candidates for start instructions that we do not currently support,
such as \texttt{LOAD_ATTR}, \texttt{LOAD_NAME}, \texttt{LOAD_GLOBAL}, \texttt{LOAD_DEREF}.
In consequence, expanding the abstract interpreter to cover more cases, i.e., more instructions and
more types, will improve performance even further.
\textsf{Spectralnorm} performs best, because our abstract interpreter finds that all of the
instructions of its most frequently executed function (\texttt{eval_A}) can be optimized.

Finally, we were surprised about the performance comparison with PyPy.
First, it is striking that we outperform PyPy 1.9 on the \textsf{spectralnorm} benchmark.
Since we include start-up and warm-up times, we decided to investigate whether this affects our
result.
We timed successive runs with higher argument numbers (1000, 1500, 2000, and 4000) and verified that
our interpreter maintains its performance advantage.
Besides this surprising result, we think that the performance improvement of our interpreter lays a
strong foundation for further optimizations.
For example, we believe that implementing additional instruction-dispatch based optimizations, such
as superinstructions~\cite{proebsting+95,ertl.gregg+03} or selective
inlining~\cite{piumarta.riccardi+98}, should have a substantial performance impact.

Second, we report that the interpreter data from~\autoref{ss:interpreter-data} compares favorably with
PyPy, too.
Using \texttt{sloccount} on the \texttt{pypy} directory on branch \texttt{version-1.9} gives the
following results.
For the \texttt{interpreter} directory, \texttt{sloccount} computes 25,991, and for the \texttt{jit}
directory 83,435 lines of Python code.
The reduction between the 100kLOC of PyPy and the 6.5kLOC of \mlq{} is by a factor of almost
17$\times$.
This is a testament to the ease-of-implementation property of interpreters, and also of purely
interpretative optimizations in general.

\section{Related Work}\label{s:related-work}

\paragraph{Partial Evaluation}
In 1996, Leone and Lee~\cite{lee.leone+96} present their implementation of an optimizing ML
compiler that relies on run-time feedback.
Interestingly, they mention the basic idea for our system:
\begin{quote}
  It is possible to pre-compile several alternative templates for the same code sequence and choose
  between them at run time, but to our knowledge this has never been attempted in practice.
\end{quote}
Substituting ``interpreter instructions''---or derivatives, as we frequently refer to them---for
the term ``templates'' in the quote, reveals the striking similarity.
In addition, both approaches leverage the compiler back-end of the ahead-of-time compiler assembling
the run-time system---in our case the interpreter.
This approach therefore automatically supports all target architectures of the base-compiler and
hence there is no need for building a custom back-end.

In similar vein to Leone and Lee, other researchers addressed the prohibitive latency requirements
of dynamic
compilation~\cite{consel.noel+96,engler.etal+96,grant.etal+97,chambers+02,philipose.etal+02} by
leveraging ideas from partial evaluation.
While we take inspiration from these prior results, we address the latency problem superimposed
by dynamic code generation by \emph{avoiding} it altogether.
Instead, we speculate on the likelihood of the interpreter using certain kinds of types and derive
optimized instructions for them.
At run-time, we rely on our novel procedure of concerting these optimized derivatives via abstract
interpretation driven multi-level quickening.
That being said, since these approaches are orthogonal, we believe that there are further
advancements to be had by combining these approaches.
For example `C, or the recently introduced Terra/Lua~\cite{devito.etal+13}, could be used to either
stage the optimized derivatives inside of the interpreter source code, or generate the necessary
derivatives at run-time, thereby eliminating the speculation part.

The initial optimization potential of partial evaluation applied to interpreters goes back to
Futamura in 1971~\cite{futamura+71}.
But, prior work has repeatedly revisited this specific problem.
In particular, Thibault et al.~\cite{thibault.etal+00} analyze the performance potential of
partially evaluated interpreters and report a speedup of up to four times for bytecode
interpreters.
This result is intimately related to our work, in particular since they note that partial evaluation
primarily targets instruction dispatch when optimizing interpreters---similar to the first Futamura
projection.
In 2009, Brunthaler established that instruction dispatch is not a major performance
bottleneck for our class of interpreters~\cite{brunthaler+09}.
Instead, our approach targets known bottlenecks in instruction implementation: dynamic typing,
reference counting operations, and modifying boxed value representations.

Gl{\"u}ck and J{\o}rgensen also connect interpreters with partial
evaluation~\cite{gluck.jorgensen+94}, but as a means to optimize results obtained by applying
partial evaluation.
Our technique should achieve similar results, but since it is speculative in nature, it does not
need information of the actual program P that is interpreted, which is also a difference between our
work and Thibault et al.~\cite{thibault.etal+00}.

\paragraph{Interpreter Optimization}
The most closely related work in optimizing high-level interpreters is due to
Brunthaler~\cite{brunthaler+10a,brunthaler+10b}.
In fact, the first-level quickening step to capture type feedback goes back to the discovery by
Brunthaler, and we have compared his publicly available system against our new technique.
In addition to the second-level quickening that targets the overheads incurred by using boxed object
representations, we also describe a principled approach to using partial evaluation for deriving
instructions.

\paragraph{Just-in-time compilers}
Type feedback has a long and successful history in just-in-time compilation.
In 1994, H\"{o}lzle and Ungar~\cite{holzle.ungar+94,holzle+94} discuss how the compiler uses type
feedback to inline frequently dispatched calls in a subsequent compilation run.
This reduces function call overhead and leads to a speedup by up to a factor of 1.7.
In general, subsequent research gave rise to adaptive optimization in just-in-time
compilers~\cite{aycock+03}.
Our approach is similar, except that we use type feedback for optimizing the interpreter.

In 2012, there has been work on ``repurposed JIT compilers,'' or RJITs, which take an existing
just-in-time compiler for a statically typed programming language and add support for a dynamically
typed programming language on top~\cite{ishizaki.etal+12,castanos.etal+12}.
This approach is interesting, because it tries to leverage an existing just-in-time compilation
infrastructure to enable efficient execution of higher abstraction-level programming
languages---similar to what has been described earlier in 2009 by Bolz et
al.~\cite{bolz.etal+09} and Yermolovich et al.~\cite{yermolovich.etal+09}, but more
invasive.
Unfortunately, the RJIT work is unaware of recent advances in optimizing interpreters, and therefore
misses some important optimization opportunities available to a repurposed just-in-time compiler.
W\"{u}rthinger et al.~\cite{wurthinger.etal+12} found that obtaining information from the
interpreter has substantial potential to optimize JIT compilation, and we anticipate that this is
going to have major impact on the future of dynamic language implementation.

Regarding traditional just-in-time compilers, Python nowadays only has one mature project:
PyPy~\cite{rigo.pedroni+06}.
PyPy follows a trace-based JIT compilation strategy and achieves substantial speedups over
standard CPython.
However, PyPy has downsides, too: because its internals differ from CPython, it is not compatible
with many third party modules written in C.
Our comparison to PyPy finds that it is a much more sophisticated system offering class-leading
performance on some of our benchmarks.
Surprisingly, we find that our technique outperforms PyPy by up to 14\% on the \textsf{spectralnorm}
benchmark, and requires substantially less implementation effort.


\paragraph{Miscellaneous}
Prior research addressed the importance of directly operating on unboxed
data~\cite{peytonjones.launchbury+91,leroy+92}.
There are certain similarities, e.g., Leroy's use of the \texttt{wrap} and \texttt{unwrap} operators
are related to our (un-)boxing functions, and there exist similar concerns in how to represent bits
in a uniform fashion.
The primary difference to the present work is that we apply this to a different language, Python,
which has a different sets of constraints and is dynamically typed.

In 1998, Shields et al.~\cite{shields.etal+98} address overhead in dynamic typing via staged type
inference.
This is an interesting approach, but it is unclear if or how efficient this technique scales to
Python-like languages.
Our technique is much simpler, but we believe it could very well benefit of a staged inference
step.

\section{Conclusions}\label{s:conclusions}

We present a general theory and framework to optimize interpreters for high-level languages such as
JavaScript, Python, and Ruby.
Traditional optimization techniques such as ahead-of-time compilation and partial evaluation only
have limited success in optimizing the performance of these languages.
This is why implementers usually resort to the expensive implementation of dynamic
compilers---evidenced by the substantial industry efforts on optimizing JavaScript.
Our technique preserves interpreter characteristics, such as portability and ease of implementation,
while at the same time enabling substantial performance speedups.

This important speedup is enabled by peeling off layers of redundant complexity that interpreters
conservatively re-execute instead of capitalizing on the ``dynamic locality of type usage''---almost
three decades after Deutsch and Schiffman described how to leverage this locality for great benefit.
We capitalize on the observed locality by speculatively staging optimized interpreter instruction
derivatives and concerting them at run-time.

First, we describe how speculation allows us to decouple the partial evaluation from any concrete
program.
This enables a principled approach to deriving the implementation of optimized interpreter
instruction derivatives by speculating on types the interpreter will encounter with a high
likelihood.

Second, we present a new technique of concerting optimized interpreter instructions at run-time.
At the core, we use a multi-level quickening technique that enables us to optimize untyped
instructions operating on boxed objects down to typed instructions operating on native machine
data.

From a practical perspective, our implementation and evaluation of the Python interpreter confirms
that there is a huge untapped performance potential waiting to be set free.
Regarding the implementation, we were surprised how easy it was to provide optimized instruction
derivatives even without automated support by partial evaluation.
The evaluation indicates that our technique is competitive with a dynamic compiler
w.r.t.~performance \emph{and} implementation effort: besides the speedups by a factor of up
to \namasteMaxSU{}, we report a reduction in implementation effort by about~17$\times$.

\bibliographystyle{abbrvnat}

\end{document}